\documentclass{aastex}
\usepackage{spr-astr-addons, amsmath,amssymb}
\usepackage{bm}
\usepackage{url}

\RequirePackage{color}

\newcommand{\emaila}

\begin{document}

\title{Turbulence induced additional deceleration in
  relativistic shock
  wave propagation: implications for gamma-ray burst}
\shorttitle{Article preparation guidelines}
\shortauthors{Short Author}

\author{Xue-Wen Liu}
\emaila{astrolxw@gmail.com}
\affil{Department of Physics, Sichuan University, Peoples
  Republic of China 610065}

\begin{abstract}
  The late afterglow of gamma-ray burst is believed to be
  due to progressive deceleration of the forward shock
  wave driven by the gamma-ray burst ejecta propagating in the
  interstellar medium. We study the dynamic effect of
  interstellar turbulence on
  shock wave propagation. It is shown that the shock wave
  decelerates more quickly than previously assumed without the
  turbulence. As an observational consequence, an earlier jet break will appear in
  the light curve of the forward shock wave. The scatter of
  the jet-corrected energy release for gamma-ray burst,
  inferred from the jet-break, may be partly due to the
  physical uncertainties in the turbulence/shock wave
  interaction. This uncertainties also exist in two shell
  collisions in the well-known internal shock model proposed
  for gamma-ray burst prompt emission. The large scatters of
  known luminosity relations of gamma-ray burst may be
  intrinsic and thus gamma-ray burst is not a good standard
  candle. We also discuss the other implications.

\end{abstract}


\section{Introduction}\label{s:intro}
Gamma-ray burst (GRB) is the most explosive event in the
universe. The standard picture for GRB is the relativistic
fireball shock model \citep{Pacz86,Goodman86,Shemi, RM92,
  MR93, RM94}. Within such a picture, an initially hot
fireball composed of photons, electron-positron pairs, and a
small amount of baryons expands outward because of the large
optical depth, converts most of its thermal energy into the
bulk kinetic energy of the baryons to form a relativistic
cold shell; the expanding shells interact with each other
and with the surrounding medium, causing their kinetic energy to
be radiated in shock waves and producing the observed GRB
prompt and afterglow emissions. Within such a scenario, the
relativistic shock generates the magnetic field via Weibel
instability {\citep{Weibel59, Medvedev99}} and the
energetic electrons via first order Fermi acceleration which
cool down, most likely via synchrotron emission {\citep{M94, Tavani96}}.
Although the standard fireball model can explain the general
features of GRB: the early-time rapid temporal variability and
late-time smooth afterglow, there are some observational features
beyond the expectations of this model such as the early
X-ray plateaus, various rebrightenings and chromatic breaks {\citep{Zhang11}}.
Different extensions of the basic model are invoked to
explain the observed deviation from the model. The
extensions include the modification of the total energy of
the ejecta, the environment, the microphysics parameters and the
radiative mechanism {\citep{Zhang11}}. However, often these
are tailored on a burst by burst basis. 

Most astrophysical system, e.g., accretion disks,
solar/stellar winds, and the interstellar medium (ISM) are
in turbulent states with embedded magnetic fields that
influence almost all of their properties {\citep{Biskamp03,
    Frisch95, Goldstein95, Elmegreen04, Scalo04}}. Narayan \& Kumar
(2009) {\citep{Narayan09}} and Lazar et al. (2009)
{\citep{Lazar09}} have proposed a relativistic
turbulence model instead of the internal shock model as the
production mechanism for fast variable GRB light curves and
applied it to GRB 080319B {\citep{Kumar09}}. Zhang \&
Yan {\citep{Zhangyan11}}  have also developed a new model of GRB prompt
emission in the highly magnetized regime, namely, the
Internal-Collision-induced Magnetic Reconnection and
Turbulence model which not only carries the merits, but also
alleviate some drawbacks of the internal shock model.
The role that the magnetic fields play in acceleration,
collimation and emission production in GRB outflows remains
one of the central issues. As an alternative to microscopic Weibel
instability, a production mechanism of magnetic field by
macroscopic turbulence is proposed {\citep{Goodman07, Sironi07}} and verified by
recent three-dimensional relativistic magnetohydrodynamics
simulation {\citep{weiqun09}}. This simulation also show that the
macroscopic turbulence produces small-scale moving magnetic
clouds which is likely a site for Fermi acceleration of
charged particles. It is possible that the turbulence may
give a self-consistent picture of GRB emission, which includes the fast
variability, the acceleration of nonthermal electrons and
the amplification of magnetic fields. 

The pre-shock turbulence can be
triggered by the interaction between GRB precursors and
the interstellar medium through fluid
instabilities. Cosmic rays acceleration in relativistic
collisionless shock can also excite large-scale turbulence in
the shock upstream \citep{Nakar06}. In this
paper, we focus on how the turbulence/shock
wave interaction modify the propagation of the relativistic
shock. Non-relativistic shock wave propagation in turbulent
interplanetary plasma had been studied, the result of which
show that the solar wind plasma turbulence may put considerable
contribution in the shock wave deceleration {\citep{Chashei96}}. In the
fireball shock model, the interaction
of ultra-relativistic ejecta with the surrounding medium
drives a relativistic shock which satisfies the
Blandford-McKee (BM)
self-similar solution and totally determines the time
evolution of the afterglow emission {\citep{BM76}}. If the turbulence
exists in the upstream medium, we show that the relativistic shock
will also amplify the turbulence behind the moving shock
font, and thus decelerates more quickly than that BM
solution predicted. We derive below the evolution solution
of the relativistic shock propagating in a turbulent medium
and discuss the implications for GRBs. 

\section{Turbulence induced additional deceleration of
  relativistic shock}
We consider a relativistic GRB jet
sweeping up the surrounding medium produced by the compact
GRB progenitor with density
$\rho=Ar^{-s}$, where {A=$nm_p$ for interstellar medium (ISM)
  environment while $\rm{A=5\times 10^{11}A_{\ast}\ g\
    cm^{-1}}$ for wind environment} {\citep{Chevalier00}}.
The hydrodynamics involves a relativistic blast
wave expanding into the wind. For an ultrarelativistic,
adiabatic blast wave, Blandford \& McKee find that
\begin{equation}
  E=\frac{8\pi A \Gamma^2R^{3-s}c^2}{17-4s},
  \end{equation}
where $E$ is the isotropic energy of the blast wave (mostly
denoted by $E_{\rm iso}$) and $R$ is the shock wave
radius. In the observer's frame, there is $t=R/(2\Gamma^2c)$
and thus $\Gamma\propto t^{(s-3)/(8-2s)}$.

The turbulence in the wind will make $\Gamma$ decay more
faster. We first review the basic idea of energy transformation process from
non-relativistic shock wave to the turbulence,
then generalize it to the relativistic shock. If the
turbulence in the regions before
and behind the shock surface is of acoustic type and it
induced sound wave is incident normally from the
front on a non-relativistic shock wave, the turbulence
energy transformation coefficient is
\begin{equation}
  \chi=\frac{W_d}{W_u}=\eta^2\Big[\frac{2\gamma}{\gamma+1}\Big]M_u^2,
  \end{equation}
where $W_d$ and $W_u$ are the energy density of turbulence in
the downstream and upstream of the shock, $M_u$ is the
upstream Mach number, $\gamma$ is the ratio of specific
heats, and $\eta\simeq
1/(\gamma+\sqrt{2\gamma(\gamma-1)})$. For $\gamma=5/3$, we
have $\chi\simeq 0.1M_u^2$ which means the considerable
amplification of turbulence for strong shock waves $M_u \gg
1$.   The source of the turbulence amplification is the shock wave energy.
Chashei \& Shishov showed that the relative level of
turbulence $\delta_d={W_d}/{p_d}=\eta^2\delta_u$ is small, which means that
the turbulence behind the shock is always weak {\citep{Chashei96}}. 

For a relativistic shock, the Rankine-Hugoniot relations with
the turbulence
taking into account have the following form
\begin{eqnarray}
  &  \Gamma_u n_u \beta_u = \Gamma_d n_d \beta_d  \nonumber\\
  &  w_u \Gamma_u^2 \beta_u^2 + p_u
  + \pi_u =
  w_d\Gamma_d^2 \beta_d^2 +p_d +\pi_d
  \nonumber\\
  &  w_u \Gamma_u^2 \beta_u + H_{\rm tu} = w_d
  \Gamma_d^2 \beta_d + H_{\rm td},
\end{eqnarray}
where $\pi_{u,d}$ are the turbulence momentum fluxes and
$H_{\rm tu,td}$ are the turbulence energy fluxes.
The relation
between $\pi_d$($H_{\rm td}$) and $\pi_u$($H_{\rm tu}$) for
relativistic shock is unclear. We just know that the energy
flux can be approximated
as $H\simeq \Gamma^2\varepsilon c$, where $\varepsilon$
is the energy density of the turbulence in the comoving
frame for relativistic flow. It is convenient to define a
relative  turbulence energy transformation coefficient
$\delta_{\rm du}\equiv
(H_{\rm td}-H_{\rm tu})/e'$ for
relativistic shock, where $e'$ is
the energy density of the post-shock medium in the
comoving frame.
Due to the energy transformation from the shock to the
turbulence, the shock flow total
energy decrease. We should adopt an energy equation in the
form {\citep{Chashei96}}
\begin{equation}
  \frac{\partial e}{\partial t}+\nabla \cdot \boldsymbol{H}=0,
\end{equation}
where $e$ is the energy density of the fireball (without the
turbulence energy) in the central engine frame, $\boldsymbol{H}$ is the
turbulence energy flux. If the
turbulent terms are neglected in (3), the BM solution is
achieved. If the turbulence is included, the BM solution is
not valid any more.

We shall use the perturbation method to solve (4) under the
condition $\delta_{\rm du}\ll 1$. In the
central engine frame, we have $dR=v_sdt$, where $v_s$ is the
velocity of the shock wave. Carrying out the volume
integration of the energy equation with including the energy
losses (caused by turbulence)  on the shock surface, we get
\begin{equation}
  v_s\frac{d}{dR}\Big(\int 4\pi
  r^2edr\Big)=-(H_{\rm td}-H_{\rm tu})4\pi R^2.
  \end{equation}
Since $\int 4\pi r^2 edr=E$ for zero approximation, and
the shock is relativistic $v_s\sim c$, the
above solution can be written in the form
\begin{equation}
  \frac{dE}{dR}\approx -4\pi \delta_{\rm du} e' R^2.
  \end{equation}
In principle, $\delta_{\rm du}$
depends on the detailed process of the turbulence/shock
interaction. To obtain an analytical solution, we assume
$\delta_{\rm du}={\rm const}$. 

$\bf{ISM\ environment\ (s=0).}$ The post-shock energy density is
$e'\simeq 4\Gamma^2n m_p c^2$ and the total energy of the
fireball is $E=8\pi nm_p\Gamma^2R^3c^2/17$. Then the solution of
(6) is 
\begin{equation}
  \Gamma\propto R^{-\frac{3}{2}}{\rm
    exp}\Big({-17\int_{R_0}^R \frac{\delta_{\rm du}}{r}dr}\Big),
  \end{equation}
where $R_0$ is the initial distance of the shock occurs.
For the case $\delta_{\rm du}=$ const, we have $\Gamma\propto
R^{-3/2-17\delta_{\rm du}}$. Using the relation $R=2\Gamma^2ct$,
the evolution solution of relativistic shock is
$\Gamma\propto t^{-3/8-\delta_{\rm du}/(16/17+\delta_{\rm
    du}/8)}\equiv t^{-3/8-f_{\rm dec}}$.

$\bf{WIND\ environment\ (s=2).}$ The post-shock energy density is
$e'=4\Gamma^2AR^{-2}c^2$ and the total energy is $E=8\pi
A\Gamma^2Rc^2/9$. We get $\Gamma\propto
R^{-1/2-9\delta_d}$ and thus the evolution solution is
$\Gamma\propto t^{-1/4-\delta_{\rm du}/(4/9+4\delta_{\rm
    du})}\equiv t^{-1/4-f_{\rm dec}}$.

Additional deceleration factors are involved as compared with
the BM solutions for ISM and wind environment respectively.
We note that the coefficient $\delta_{\rm du}$ in
fact includes the energy transformation process between the
shock wave and the turbulence, which may be in principle
defined more precisely. By defining it, our results can also
be applicable for MHD turbulence.

\section{Implications and discussions}
In this work, we have
demonstrated that the relativistic blast wave involving the
turbulence in the upstream decelerates more quickly than
what BM solution predicted based on the condition that the relative
turbulence energy transformation is small. This dynamical
effect changes temporal behavior of the peak frequencies and
thus the slope of the light curves. We restrain our
discussions for ISM case. The position of
the peak of the spectrum $F_\nu$ varies as $\nu_c\propto
t^{-1/2+4f_{\rm dec}}$ in fast
cooling while $\nu_m\propto t^{-3/2-4f_{\rm dec}}$ in slow
cooling. The main results are summarized in Tab. 1. Emission
for a forward shock predicts definite relation between the
spectral and temporal indexes, $F_\nu\propto
t^{-\alpha}\nu^{-\beta}$ {\citep{Rees98}}. Observed afterglow
typically do not comply with this prediction
{\citep{Racusin09}}. The deviation of the observed closure relation from
the external shock afterglow model in those bursts
such as GRB050413,
GRB0607A and GRB061202 may be caused by the upstream
turbulence {\citep{enwei07}}.

The outflows of GRB are believed to be ultrarelativistic
jets. Since the outflow decelerates more quickly, a jet-break
in afterglow will appear earlier when $\Gamma\sim 1/\theta_j$,
where $\theta_j$ is the jet opening angle. Using
$\Gamma\propto R^{-3/2-17\delta_{\rm du}}$ and
$\Gamma_0^2R_0^3=17E_{\rm iso}/8\pi nm_pc^2$ ($\Gamma_0$ is
the initial Lorentz factor of the jet), we get the jet-break
time
\begin{equation}
  t_j\sim1.4\Big(\frac{E_{\rm
      iso,53}}{n_0}\Big)^\frac{1}{3}\theta_{j,-1}^\frac{8}{3}(\Gamma_0\theta_j)^
  {-\frac{68\delta_{\rm du}}{9+102\delta_{\rm du}}} \ {\rm days},
  \end{equation}
where $\theta_j=0.1\theta_{j,-1}$, $E_{\rm iso}=10^{53}E_{\rm
  iso,53}\ {\rm erg}$ and $n=1n_0\ {\rm cm^{-3}}$. (For a
GRB located at redshift $z$, the observed time should be
increased by a factor $1+z$.) The jet-break time is earlier
by a factor of the last term in (8) than that of the
adiabatic blast wave predicted. For $\Gamma_0=1000$,
$\theta_j=0.1$ and $\delta_{\rm du}=0.1$, the value of the factor is about
$0.2$ which means that the previous jet opening angle is
notably overestimated. The jet-corrected energy $E_\gamma\simeq
\frac{1}{2}\theta_j^2E_{\rm iso}$ should be potentially
re-evaluated by (8). If the turbulence/shock interaction is
universal, i.e., the same value of $\delta_{\rm du}$ for all
GRBs, the three parameters ($t_j,E_{\rm iso}, \Gamma_0$) should be
statistically correlated. Once obtain the correlation, we
can evaluate $\theta_j$ and $E_\gamma$. While it's not easy
to identify the jet-break \citep{Liang08, Racusin09}. Because in most GRBs, the optical
and X-ray breaks are chromatic which is contradict with the forward
shock model. A natural but not affirmative way to remove the
contradiction is to
invoke a different origin of X-ray emission: dust scattering
{\citep{Shao07}} or photosphere emission of the engine
activity {\citep{Wu11}}. So the late optical observation is
crucial to identify the jet-break. Secondly, the
detection of very early optical
afterglow peak can provide a direct measurement of the
initial Lorentz factor {\citep{Vergani07}}. There are many
optical afterglows without detailed X-ray coverage before
{\sl Swift} era. Now, that {\sl Swift}-XRT provides
impressive X-ray light curves, there are too few optical
light curves. The
Chinese-French mission SVOM is a multi-wavelength GRB
observatory scheduled to launch in 2014-2015
{\citep{Paul11}}. Its operation window overlapping with that of
{\sl Fermi} and {\sl Swift} will shade light on this problem.

\begin{table}
\caption { Temporal index $\alpha$ and spectral index $\beta$
  in afterglow for ISM case, the convention $F_\nu\propto
  t^{-\alpha}\nu^{-\beta}$ is adopted. The power law index
  of the electron distribution is $p>2$.}
\def\frh{}
\def\dn{2b(4-k)}
\def\dthin{2(7-2k+s)}
\begin{tabular}{cccccccccc}
\hline \hline
&$\alpha$  &$\beta$ &$\alpha(\beta)$\cr

\hline
${\rm slow \ cooling}$\cr
\hline
$\nu<\nu_m$   &$-\frac{1}{2}-\frac{4}{3}f_{\rm dec}$
              &$-\frac{1}{3}$
              &$\alpha=\frac{3}{2}\beta+4f_{\rm dec}\beta$\cr

\hline
$\nu_m<\nu<\nu_c$    &$\frac{3(p-1)}{4}+2f_{\rm dec}(p-1)$ 
                     &$\frac{p-1}{2}$    
                     &$\alpha=\frac{3}{2}\beta+4f_{\rm dec}\beta$               \cr

\hline
$\nu_c<\nu$          &  $\frac{3p-2}{4}+2f_{\rm dec}(p-2)$ 
                     &  $\frac{p}{2}$   
                     &
                     $\alpha=\frac{3\beta-1}{2}+\frac{4f_{\rm
                       dec}(p-2)}{p}\beta$           \cr  
\hline
${\rm fast \ cooling}$ \cr
\hline 
$\nu<\nu_c$  &$-\frac{1}{6}+\frac{4}{3}f_{\rm dec}$
             &$-\frac{1}{3}$    
             &$\alpha=\frac{\beta}{2}-4f_{\rm dec}\beta$\cr

\hline
$\nu_m<\nu<\nu_c$  &$\frac{1}{4}-2f_{\rm dec}$
                   &$\frac{1}{2}$
                   &$\alpha=\frac{\beta}{2}-4f_{\rm dec}\beta$               \cr

\hline
$\nu_c<\nu$      &  $\frac{3p-2}{4}+2f_{\rm dec}(p-2)$
                 &  $\frac{p}{2}$   
                 &  $\alpha=\frac{3\beta-1}{2}+\frac{4f_{\rm
                   dec}(p-2)}{p}\beta$           \cr  
\hline\hline
\end{tabular}
\end{table}

The turbulence/shock interaction make some shock energy
stored in turbulent state, which lead to some
uncertainties and differences between the observational and
real values of $t_j$, $E_{\rm iso}$, $L_{\rm iso}$ (isotropic
luminosity) and $E_{\rm peak}$ (peak energy of $\nu F_\nu$
spectrum). A number of relations involving these parameters
were proposed (see, e.g., {\citep{Frail01, Amati02,
    Ghirlanda04, Yonetoku04, Liang05}}), which lead to
identification of a ``GRB standard candel'' for cosmology
information complementing that derived from SNe. This
requires that the energy and the luminosity are precisely
estimated from observational quantities. But the significant
dispersions of these correlations prevent GRB as a good
standard candle, which may be intrinsic due to the lack of
the full knowledge of the turbulence/shock interaction.

Those quantities related to the shock will be effected by the
turbulence, so do the correlations between these
quantities. Although current understanding of the
turbulence/shock interaction is limited, posing a challenge
to accurate prediction of this highly non-linear phenomenon,
the results show the potential ability of the turbulence to
solve some problems in GRBs. The results of this letter are
also applicable for SNe, AGN and microquasar.

Xue-Wen Liu thank Jin-song Zhao for valuable
discussions. This work is supported by
the National Natural Science Foundation of China (grant
11003014/A0303).

\end{document}